\pdfoutput=1

\documentclass[preprint,12pt]{elsarticle}
\usepackage[utf8]{inputenc}
\usepackage{hyperref}

\usepackage{amsmath,amssymb}
\usepackage{amsthm}
\usepackage{latexsym}
\newtheorem{theo}{Theorem}
\newtheorem{lemme}{Lemma}
\newtheorem{defi}{Definition}
\newtheorem{coro}{Corollary}
\newtheorem{prop}{Proposition}
\theoremstyle{definition}
 \newtheorem{example}{Example}

\newenvironment{myproof}{ 
\noindent {\bf Proof} \noindent} {\hfill $\Box$\vskip 5mm} 

\journal{Discrete Applied Mathematics}

\begin{document}
\begin{frontmatter}



\title{Finite Confluences and Closed Pattern Mining }


\author[]{Henry Soldano}

\address[a]{ L.I.P.N UMR-CNRS 7030, Université  Sorbonne Paris Nord,\\
F-93430, Villetaneuse, France
}
\address[b]{Atelier de BioInformatique, ISYEB - UMR 7205 CNRS MNHN UPMC EPHE, Museum d'Histoire Naturelle, F-75005, Paris, France
}
\address[c]{ NukkAI, F-75013, Paris, France
}
\begin{abstract}
The purpose of this article is to propose and investigate a partial order structure weaker than the lattice structure and which have nice properties regarding closure operators.   We extend accordingly closed pattern mining and   formal concept analysis to such structures we further call confluences.  The primary motivation for investigating these structures is that it allows to reduce a lattice to a part whose elements are  connected, as in  some graph, still preserving a useful characterization  of closure operators. Our investigation also considers how reducing one of the lattice involved in a Galois connection affects the structure of the closure operators ranges.  When extending this way formal concept analysis we will focus on the intensional space, i.e. in reducing the pattern language, while 
recent investigations rather explored the reduction of the extensional space to connected elements. 

\end{abstract}


\end{frontmatter}
\newpage
\tableofcontents
\newpage

\section{Introduction}

Until recently, searching for \emph{closed} patterns or patterns when exploring data was restricted to lattices as pattern languages. A pattern  in some language $\ L$ is said closed whenever it can be obtained by applying a \emph{closure operator} to some pattern. This subject has been thoroughly explored both  from a mathematical and algorithmical point of view as well in formal concept analysis (further referred to as FCA)  \cite{Gan99}, Galois analysis \cite{Caspard2003xi,Diday2003rm} and, more recently, in data mining \cite{pasquier99}. Most of this work considers \emph{support-closed} patterns. In this case, we  consider a set of objects $O$ and a  pattern may \emph{occur} or not in each object.  The \emph{support set} of a pattern is then the subset of objects in which the pattern occurs. The language is a lattice with respect  to  a \emph{general-to-specific} ordering and each object is described by  a pattern.  
Patterns that cannot be specialized without reducing their support set are said \emph{support-closed}. Clearly, there is some redundancy in enumerating all patterns when we are concerned with properties relative to their occurrences, such as \emph{frequency}, and it is interesting to only consider support-closed patterns. In a lattice,
support-closed patterns may be efficiently searched for
because there exists a closure operator on the lattice that returns, as the closure $c=f(t)$ of some  pattern $t$, the unique most specific support-closed pattern sharing the same support set $\mathrm{ext}(t)$ as $t$, i.e. the most specific element of the equivalence class of patterns whose support set is $\mathrm{ext}(t)$.  In such a case, a Galois connection relates the intensional space, i.e. the pattern language, and the extensional space, i.e. the powerset of objects, and a closure operator is defined on each space. The corresponding  set of  closed elements pairs is then a lattice, we refer to, in general, as a Galois lattice, and more precisely as a concept lattice  when the connection relates an intensional  space and an extensional space. In this case,  the elements of the $( \mathrm{ext}(c), c)$  pair represents respectively the \emph{extent} and the \emph{intent} of a concept.  Furthermore, by rewriting inclusion of support sets as  valid implications on the object set, we may define, in various ways, an implication basis such that all valid implications are generated from the implication basis using a set of logical rules \cite{duquenne86,Gan99,Caspard2003xi,Bertet:2010uq,pasquier99}. Finally, recent work have shown that applying an interior operator  to either  the extensional space $X=2^O$ \cite{Soldano:2011fk} or to the \emph{intensional space}  as proposed in \cite{ganter01pattern} or both \cite{pernelle2002oe} led to a new Galois connections, and to a concept lattice, smaller than the original one, we further refer to as an abstract concept lattice. As a dual of a closure operator, the range of  an interior   operator defined on a lattice $T$ is a subset closed under join and will be further called an \emph{abstraction of $T$}. 
 The most investigated pattern language, in particular in FCA, is the 
power set $2^I$ of some attribute set $I$.
However various  work have shown that the whole methodology was preserved  when considering as a pattern language any complete  lattice,  as for instance a finite lattice of intervals.  Direct application of the standard FCA methodology  for such settings is always possible by embedding the pattern language in the power set of some attribute set, therefore allowing to have generic algorithms, but they can also be addressed directly as far as the language is a finite lattice 
\cite{liquiere98,ganter01pattern,POLAILLONri,pernelle2002oe}.   

 Recently, pattern mining has gone beyond this general framework in two directions. 
 First, various mining problems have been investigated that come down to searching for closed patterns which cannot be considered, strictly speaking, as support-closed patterns, such as  convex hulls of subsets of a given set of points, or sequential patterns with wild-cards. Solving the problem then mainly means  defining and building the corresponding closure operator \cite{Arimura2009vn}. To characterize such closure operators,  the authors make direct use  of the well-known theorem, on which relies also FCA,  stating that in a finite lattice $T$ there is an equivalence  between the families closed under the \emph{meet} operator 
 and the closure operators on $T$ (see for instance \cite{Blyth2005}).  This means there is a practical  interest in closure operators that goes beyond  establishing a relationship between two partial orders. 
 Second, various mining problems have been addressed in which the pattern language is  not a lattice,  in particular  problems where closed patterns are support-closed patterns with respect to some dataset of objects. A  framework has been proposed for that purpose in which the language is a family $F$ included in a host lattice $2^I$. 
 For instance, consider the set of the subgraphs generated by subsets of the edge set $I$ of a  graph $G$. Searching for support-closed patterns w.r.t. some object set can be performed as a standard $2^I$ lattice  mining problem. 
However if we want to consider as a language the family $F$ of connected subgraphs of $G$, represented by their edge subset, then $F$ is not a lattice\footnotemark,
 \footnotetext{ the intersection of  two such connected subgraphs is not necessarily connected}
 still  there is a closure operator that applied to some connected edge subset $x$ return  the unique support-closed connected edge subset greater than $x$.
  This means that we can use the  same kind of algorithm that specializes a closed pattern, computes the support  of the new pattern and  closes it in the same way we do in the lattice  mining case.  In their paper \cite{Boley:2010fk} M. Boley and coauthors state in particular the necessary and sufficient conditions that  the family $F$ of subsets of $I$ has to fulfill  in order to guarantee that, whatever the  dataset $O$ of objects is\footnotemark, there exists a closure operator to compute support-closed patterns. The present article found its motivation in the latter work and investigates to what extent standard results about closure operators  in lattices  may be extended in such a way that formal concept analysis itself is extended while preserving its ability to preserve order structures.
   \footnotetext{with some mild restriction we discuss further}

We define and investigate a partial order structure, we further call a confluence, weaker than the lattice structure, which have nice properties regarding closure operators  and  extend accordingly  formal concept analysis.  The primary motivation for investigating confluences  is that they allow  in particular   to reduce a lattice to a part whose elements are  connected elements with respect to some graph, still preserving a useful characterization  of closure operators, so widening their use in pattern mining and FCA.  Our investigation goes then  on   how reducing one of the lattice involved in a Galois connection affects the structure of the closure operator ranges, and then investigates more specifically how FCA can be extended this way, focussing on the intensional space, i.e. in reducing the pattern language.

First, we characterize closure operators and their range in a confluence. A confluence is characterized as a finite partial order  $F$ with possibly several minimal elements $m$ and such that each of its principal up-set $F^m$ is a lattice. We say then that in $F$ it exists  \emph{local meet operators} and  show that in $F$ the join $\lor_m$ and $\lor_{m'}$ operators of two principal up-sets $F^m$ and $F^{m'}$ return the same element when applied to elements belonging to both  up sets, and as a consequence we may define a single \emph{local join operator} $\lor_F$.  We extend then  the theorem on finite lattices mentioned above to confluences and obtain  in particular that the set $f[F]$ of closed elements of a confluence $F$  also is  a confluence.  We investigate then  \emph{confluences of a lattice $T$} or \emph{lattice confluences}.  A confluence $F$ of $T$ is defined as a subset of $T$ which is a confluence whose local join operator $\lor _F$ is the join operator of $T$. We see then that confluences  generalize abstractions as defined above.
When considering a powerset as the host lattice  $T$, a confluence is close to, but differs to the notion of confluent set systems as defined in \cite{Boley:2010fk} (see Section \ref{confluenceDeGalois}).
We  investigate then what happens when, starting from a Galois connection between two lattices $L$ and $X$, we reduce $L$ to be a confluence   of it: we obtain then a Galois confluence, i.e. a set of pairs of closed elements,  so generalizing the Galois lattice definition.  Second, we apply these results to extend the notions of concept lattice and abstract concept lattices, here in reducing the pattern language, to
 concept confluences and related implication bases. Finally we propose an efficient algorithm, extending the algorithm proposed in \cite{Boley:2010fk}, to compute the set of abstract closed patterns of a strongly accessible subconfluence of $2^I$.  
 
The article is organized as follows. In Section \ref{sous-ensembles} we discuss closure operator on lattices before defining  and investigating in Section \ref{confluences} confluences and confluences. In Section \ref{FromGaloisLattice} we investigate the connection between a lattice and a confluence  deriving from the reduction of the second lattice to one of its confluence. In Section \ref{formalConceptAnalysis} we briefly summarize concept lattices and abstract concept lattices as a methodology for analyzing the unknown structure of the data. In Section \ref{confluenceDeGalois} we investigate  how formal concept analysis is  extended when reducing the pattern language  to a confluence, and introduce concept confluences. Section \ref{algorithmique} discuss the algorithm mentioned above that lists concepts of a confluence. Finally we discuss this work in Section \ref{discussion} before concluding.

Lattice confluences and confluences were first defined and investigated in \cite{Soldano:2014aa} together with intensional concept confluences and related implications. Part of Section \ref{FromGaloisLattice} also benefits from the investigation on extensional concept confluences in \cite{Soldano2015fk}. Overall,  Section \ref{confluences} and Section \ref{FromGaloisLattice} represents a complete presentation of  the necessary definition and results to reduce either the intensional or the extensional space in FCA to confluences. Algorithm 1 of Section \ref{algorithmique} is presented here for the first time.

\section{ Closure operators on  a lattice   }\label{sous-ensembles}

\subsection{Closure and interior operators}

An ordered set is a set $E$ on which  a partial order   $\leq$ is defined. A $\wedge$-semilattice is an ordered set in which any elements pair $x$, $y$  has a greatest lower bound $x \land y$, also called their \emph{meet}. Dually in a $\vee$-semilattice  any pair $x$, $y$ has a least upper bound $x \lor y$, also called their \emph{join}. A lattice is both a $\land$-semilattice  and a $\lor$-semilattice. 

All ordered sets mentioned in this article are finite. As a consequence,  all lattices are  \emph{complete}, i.e. any part  $C$   has a least upper bound and a greatest lower bound,  $\wedge$-semilattices with a maximum (a top element $\top$)   and  $\vee$-semilattices with a minimum( a bottom element  $\bot$) are lattices. Furthermore all lattices have both  a top element and a bottom element.

We say that a subset $C$ of a lattice $T$  \emph{is  closed under meet} whenever the  greatest lower bound   $\bigwedge_{c \in C'} c$ of any part $C'\subseteq C$ also belongs to $C$. This definition also applies to the empty set and therefore   $ \top={\bigwedge}_{\substack{ \emptyset}} c $ also belongs to $C$.
As a consequence,  $C$ being  finite   is a lattice. Dually, a part $A$ of a lattice  $T$ is said \emph{closed under join}  whenever the lowest upper bound $\bigvee_{a \in A'} a$ of any part  $A' \subseteq A$   also belongs to $A$ and  therefore $\bot$ also belongs to $A$.  
In what follows a part of a lattice $T$ closed under join is also  called an   \emph{abstraction} of $T$. 

\begin{example}\label{ex1}
Consider the lattice  $T= 2^{abcd}$ with the inclusion order $\subseteq$ and  set theoretic intersection $\cap$ (respectively set theoretic union $\cup$) as  meet  (respectively join) operators.
\begin{itemize}
\item $A=\{\emptyset, a, c, abc\}$ is a lattice but is not closed under meet (as $abcd \not \in A$)   and is not closed  under join (as $a \cup c=ac \not \in A$)

\item $C=\{a, ab, ac, abcd\}$ is closed under  meet but not under join (as $\emptyset$ does not belong to $C$). 
\end{itemize}

\end{example}

Let $E$ be an ordered set, we will denote by   $E^t =\{x \in E~|~x \geq t\}$ its  principal  \emph{up sets} and  $E_t =\{x \in E~|~x \leq t\}$ its principal down sets with respect to a given element $t$. 
We first recall definitions of closure and dual closure operators:
\begin{defi}
Let $E$ be an ordered set and $f: E\rightarrow E$ be an map such that for any $x,y \in E$,  $f$ is monotone, i.e. $x \leq y \implies f(x) \leq f(y)$ and idempotent, i.e. $f(f(x)=f(x)$, then:
\begin{itemize}
\item if  $f$ is  extensive, i.e. $f(x) \geq x$, $f$ is called a closure operator
\item if $f$ is  intensive, i.e. $f(x) \leq x$, $f$ is called  a dual closure operator.
\end{itemize}
 In the first case, an element such that $x=f(x)$ is called  a closed element.
\end{defi}
Results regarding closure operators have dual results regarding dual closure operators. Such dual closure operators are also called interior operators or kernel operator or also projections depending of the context. In the remaining of this article  we call them \emph{ interior operators}.
We define hereunder a \emph{closure subset} of an ordered set $E$ as the range  of a closure operator, and give  the dual definition of an\emph{ interior subset}. 

\begin{defi}[T.S. Blyth \cite{Blyth2005}]
Let $E$ be an ordered set, 
\begin{itemize}
\item a subset   $C$ is called a closure subset if there is a closure operator $f: E \rightarrow E$ such that $C= f[E]$.
\item  a subset  $A$ is called an interior subset   if there is an interior operator $p: E \rightarrow E$ such that $A= p[E]$.
\end{itemize}
\end{defi}
The following result gives  a general characterization of closure subsets of an ordered set:
\begin{prop}[T.S. Blyth \cite{Blyth2005}]\label{propFond}
A subset $C$ of  an ordered set $E$ is a closure subset of $E$ if and only if for every $x \in E$ the set $C \cap E^x$ has a bottom element $x_*$. The closure $f: E \rightarrow E$ is then unique and defined as $f(x)  =x_*$.
\end{prop}
Dually $A$ is an interior subset of $E$ if and only if for every $x \in E$ the set $A \cap E_x$ has a top element. Interior operators have the interesting property to produce closure operators when composed with closure operators:

\begin{prop}[See proof \ref{Preuves}]\label{pRondf}
Let $E$ be an ordered set,  $f$ be a closure operator on $E$ and $p$ be an interior operator on $E$, then $p\circ f $ is a closure operator on $p[E]$

\end{prop}

\subsection{Closure subsets and interior subsets of a lattice}

A well know results states then  that the closure subsets of a complete lattice are the subsets closed  under the meet operator $\wedge$  \cite{Gan99}: 

\begin{prop}\label{proFond}
Let  $T$ be a  lattice, a subset  $C$  of $T$ is a closure subset if and only if  $C$ is closed under meet. The closure $f: T \rightarrow T$ is then unique and defined as $f(x)  ={\bigwedge}_{c \in C \cap T^x} c$.

\end{prop}

The dual result states that  a subset $A$ of a lattice $T$ is an  interior subset  whenever $A$ is closed under join. The interior operator $p: T \rightarrow T$ is then defined as  $p(x)  ={\bigvee}_{a \in A \cap T_x} a$. 
Because of their role in simplifying the view of data in data analysis the set $A=p[T]$ of the interior subsets of a lattice is called an \emph{abstraction}. Technically an abstraction  is a   sub-join-semilattice with same minimum as its host  lattice.

\begin{example}\label{ex2}
Consider the lattice $T= 2^{abcd}$ of example $\ref{ex1}$
\begin{itemize}

\item $A=\{\emptyset, ab, ac, abc\}$ is an abstraction of $T$ whose associated interior operator $p$ is such that   $p(x)$ is the greatest element of $A$ which is included in  $x$. We then have, $p(a)=\emptyset$ and $p(abcd)=abc$.
\item $C=\{a, ab, ac, abcd\}$ is a closure subset    of $T$. The associated closure operator  $f$ is such that $f(x)$ is the least element in $C$ which contains $x$. We then have, $f(abc)=abcd$ and $f(ab)=ab$.
\end{itemize}

\end{example}

We deduce from Proposition  \ref{pRondf} this  straightforward corollary allowing to define a closure operator on an abstraction from closure the closure operator on its host lattice:
\begin{prop}\label{pRondfA}
Let $A$ be an abstraction of a lattice $T$,   $f$ be a closure operator on $T$ and $p$ be the interior operator associated to $A$, then $p\circ f $ is a closure operator on $A$ 

\end{prop}

This means that by reducing the lattice $T$ through an abstraction, we preserve the existence of a closure operator. This result, though then not explicitly stated,  is  the basis of works on abstract concept lattices\cite{pernelle2002oe,Soldano2011fk}. 

We are interested  now in confluences which are structures weaker than lattices in which a characterization  of  closure subsets generalizing Proposition \ref{proFond} holds.

\section{Closure operators on a confluence}\label{confluences}
\subsection{Confluence as a generalization of the lattice structure}
\begin{defi}\label{def-Preconfluence}
Let  $F$ be an ordered set such that  for any $t \in F$,   $F^t $ is a $\land$-semilattice with a greatest element $\top_t$. 
$F$ is called  a confluence, $x \land_t y$ is a local infimum or local meet, and  $\top_t$ a local top.
\end{defi}

From now on, as we consider  finite ordered sets, all confluences are finite and  in the definition $F^t $ is a lattice. 
Hereunder we give  a lemma stating that given two elements  their various  local joins   coincide whenever upper bounds exists. 
\begin{lemme}[See proof \ref{Preuves}]\label{lemmeFond}
Let $F$ be a confluence, then for any $ t$ in $F$ and $x,y \in F^t$, 
\begin{enumerate}
\item   
The join  $x \vee_t y$   is the least element of 
 $F^x \cap F^y$ and is denoted by  $x \lor_F y$. 
\item Let  $t' \geq t$ then  $F^{t'} $ is a sublattice of $F^t $ i.e. $\land_{t'}=\land_t$.

\end{enumerate}

\end{lemme}
\begin{myproof} 
\begin{enumerate}
\item 

When $z \geq t$, the elements of $F$ greater than or equal to $z$ are also the elements of $F^t$ greater than or equal to $z$,  i.e. $(F^t)^z=F^z$, and therefore, let $x, y \in F^t$, $x \vee_t y$  is the lowest element of  $(F^t)^x \cap (F^t)^y$ i.e. the lowest element of $F^x \cap F^y$.

\item  Whenever $t' \geq t$, we have that $F^{t'}= (F^t)^{t'}$ and as a principal up set of $F^t$ it is a sublattice of $F^t$.
 
\end{enumerate}
\end{myproof} 

 Furthermore, as we are in the finite case, we  only need minimal elements of $F$ to check whether $F$ is  a confluence:
 
 \begin{lemme}  \label{lemmeMin}
 $F$ is a confluence if and only if for any $m\in \mathrm{min}(F)$, $F^m$ is a lattice.  
 \end{lemme}
\begin{myproof}  {\bf Lemma  \ref{lemmeMin}}

Let $M=\mathrm{min}(F)$.  ($\Rightarrow)$ If  $F$ is a confluence, then all $F^m$ are lattices. ($\Leftarrow)$  Suppose all  $m$ in $M$ are such that $F^m$ is a lattice, and consider any $t \geq m$ and two elements  $t_1,t_2 \in F^t$, we have then that  $t_1,t_2 \in F^m$. We know that $t_1 \land_m t_2$ is the greatest lower bound of $t_1$ and $t_2$ in $F^m$. As $t$ is a lower bound of  $\{t_1,t_2\}$ and $t \in F^m$, we have that $ t_1 \wedge_m t_2 \in F^ t$. Therefore  $ t_1 \wedge_m t_2$ also is the greatest lower bound of $\{t_1,t_2\}$ in  $F^t$ and as a consequence $t_1 \wedge_t t_2$ exists  and that means that   $F^t$ is a $\wedge$-semilattice. Furthermore, $\top_m$  belongs to  $F^t$ and so $F^t$ has a maximum and therefore $F^t$ is a lattice. As for any $t\in  F$ there exists some  $m \in M$ with $t \geq m$, $F$ is a confluence. 
\end{myproof}

As a corollary of the latter   Lemma we obtain that confluences generalize lattices:
\begin{lemme}
A lattice  is a confluence with a minimum.
\end{lemme}

\subsection{Subsets closed under local meet as closure subsets }\label{sous-ensembles-locaux}
We have generalized in the previous section the definition of a lattice by  considering a partial order $F$  each up-set  $F^t$ of which  is a lattice. 
 Our purpose is now to generalize Proposition   \ref{proFond}. We start defining what we call a subset \emph{closed under local meet}. 
\begin{defi}\label{defPreconf}
A subset $C$ of a  confluence $F$ is called  closed under local meet whenever for any  element $t \in F$,  $C\cap F^t$   is closed under $\land_t$.

 \end{defi}
 \begin{lemme}
 Let $C\subseteq F$ be closed under local meet in $F$, then 
 	$C$ is a confluence with same local meet operators as $F$.
 \end{lemme}
 \begin{myproof}
 
 For any $t \in C$,  $C\cap F^t$ has $t$ as its minimal element and may be rewritten as $C^t$.  $C^t$ being closed under $\land_t$ is a lattice. As a result, $C$ is a confluence whose local meet operators are the same as those of $F$. 
 
 \end{myproof}

 \begin{example}
Let  $F=\{a,b,abc\}$.
\begin{itemize}
\item  $\{a,b, abc\}$ is closed under local meet as $\{a,abc\} $ is closed under $\land_a$ and $\{b,abc\} $ is closed under $\land_b$. 
\item  $\{a,b\}$ is not closed under local meet as $\{a\} $ is not closed under  $\land_a$:  $ \land_{a_{\emptyset}} = abc$ does not belong to   $\{a,b\}$. 
\end{itemize}
 \end{example}
 
We obtain then that  
the following theorem extends Proposition \ref{proFond} to confluences:

\begin{theo}[See proof \ref{Preuves}]\label{thFondConf}

Let $F$ be a confluence.
A subset  $C$  of  $F$ is a closure subset if and only if $C$ is closed under local meet.
The closure operator $f: F \rightarrow F$ is then defined as  $f(t)  ={{\bigwedge}_t}_{\{c \in C \cap F^t\}} c$.

\end{theo}

Note that this means that  $C=f[F]$ also is  a confluence.
As a consequence of Lemma  \ref{lemmeMin} we have that  $f(t)$ rewrite as ${{\bigwedge}_m}_{\{c \in C \cap F^t \}} c$ for any $m\in min(F)$ s.t. $t \geq m$.

\begin{figure}[!htbp]

\begin{center}

\includegraphics[width=4.5in]{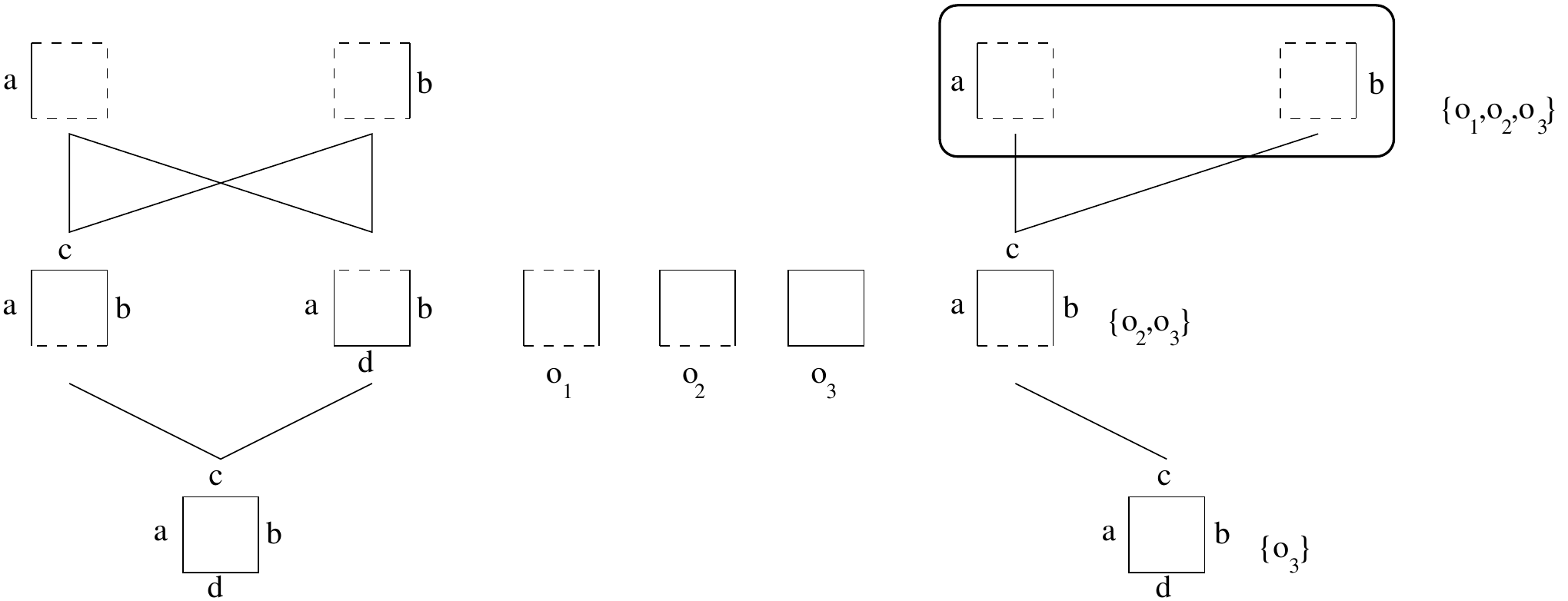}
\caption{The diagram on the left represents a family $F$ of connected subgraphs each generated by a  subset (represented by a word) of the edges $\{a,b,c,d\}$ of the original graph. The subgraphs generated by $a$ and by $b$ are the minimal elements.
The  diagram on the right represents the  concept  confluence $f[F]$ (see Section \ref{subsecExample}) with respect to the set of subgraphs $O=\{o_1, o_2, o_3\}$ represented on  the middle  part of the figure.  The closed patterns $abc$ and $abcd$ represent  the greatest connected subgraphs whose support sets  are respectively $\{o_2, o_3\}$ and $\{o_3\}$. The thick box around closed patterns $a$ and $b$ indicates that both patterns have the same support set $\{o_1,o_2,o_3\}$.}
\label{fig-confPlusObjPlusPreconf}
\end{center}
\end{figure}

 As a summary, we have a generalization of the meet operator which is the basis of most work on closed patterns in data mining, as in  formal concept analysis. 
 \begin{example}\label{connectedEdgeSubsets}
 A typical example of such a structure is the set of connected subgraphs induced by the vertices (or edges) of a given graph. We consider here a family    $F=\{a,b, abc, abd, abcd \}$, representing  connected edge subsets of a graph,  which  diagram is represented in the leftmost part of Figure \ref{fig-confPlusObjPlusPreconf}.  Here we have that $abc \land_a abd = a$ and  $abc \land_b abd = b$ i.e. there  are two maximal  lower bounds of $abc$ and $abd$ in $F$ because $ab$ does not belong to $F$.  Note that the up sets $F^a$ and $F^b$ are lattices,  and share the same join operator, which in this case is the union operator.
 \end{example}

The main motivation here is therefore to  address problems in which such connectivity constraints are appropriate. This means that we are interested in the particular case, exemplified above, of confluences included in a host lattice and sharing the same joins.
\subsection{Subconfluence of a lattice  as a generalization  of the abstraction structure}

We have previously define an abstraction $A$ of a lattice $T$ as a subset of $T$ closed under the join operator, and  also as the range $A=p[T]$ of an interior operator $p$. It was also shown that composing an interior operator $p$ with a closure operator $f$ on a lattice $T$ we obtain  a closure  operator $p\circ f$ on the abstraction $A=p[T] $.  In this section we generalize the abstraction definition by defining  a subconfluence  $F$ of a lattice $T$ as a subset of $T$  which is a confluence with same join as $T$ and is associated with a family of interior operators $\{p_t\}$.  More generally, a subconfluence is a subset of a confluence which share the same join. We will then show  that reducing the lattice to such a structure $F$  we obtain a closure operator $f_F$ on $F$ such that $f_F(t)=p_t\circ f (t)$.  

\begin{defi}\label{confluence}
Let $T$ be a lattice and $F \subseteq T$ be a confluence with as   join   $\vee_F = \vee$,   $F$ is then called a subconfluence of $T$.

\end{defi}

Proposition \ref{propEqui}, hereunder, gives two alternative definitions of subconfluences. In particular, 
the characteristic property  \ref{propEqui}.\ref{join}  is close to, but differs from, the definition of a confluent system given by M. Boley and co-authors \cite{Boley:2010fk} and that we further discuss in Sections \ref{confluenceDeGalois}  and \ref{algorithmique}.

\begin{prop}\label{propEqui}
Let $F$ be a subset of a lattice $T$, the  following statements  are equivalent: 
\begin{enumerate}
\item $F$ is a subconfluence of $T$ \label{pre}
\item For any $t \in F$, $F^T$ is a dual closure subset of $T^t$ with interior operator $p_t$ defined by $p_t(x)=\bigvee_{q \in F^t \cap T_x} q$ \label{great}
\item  For any $x,y, t$  in $F$ such that  $x$ and $y$ belong to $F^t$, $x \vee y$ belongs to $F$ \label{join}

\end{enumerate}

\end{prop}

\begin{myproof}

\ref{pre} $\Rightarrow$   \ref{great}  $F^T$  is a part of the sublattice $T^t$ which is closed under $\lor$ and following the dual of Proposition \ref{proFond} this is equivalent to say that $F^T$ is a dual closure subset of $T^T$. 

\ref{great} $\Rightarrow$   \ref{pre} as  for any $t$, $F^t$ as a dual closure subset of $T^T$ is a lattice which have  $\lor$ as join operator.

\ref{great}  is equivalent to  \ref{join} following the dual of  Proposition \ref{proFond}.

\end{myproof}

\begin{example}
Example \ref{connectedEdgeSubsets} displayed  the set $F$ of edge  subsets of a graph $G=(X,E)$  as an example of confluence. As noticed then, the join in $F$ is the set theoretic union and therefore $F$ is a subconfluence of $2^E$.

\end{example}

We obtain straightforwardly  that subconfluences generalize abstractions:

\begin{lemme}
An abstraction of a lattice $T$  is a subconfluence of $T$  with a minimum which is $\bot_T$

\end{lemme}
Again, by considering  abstractions of a lattice $T$ as  subconfluences containing the minimum of $T$, we find  a previous result, here the dual of Proposition \ref{proFond}, that characterizes abstractions as ranges of interior operators on lattices: whenever the subconfluence $F$ is an abstraction, there is only one interior operator to consider, namely $p_{\bot}$ which is such that $p_{\bot}[X]=F$. It will prove useful to note that these  interior operators are ordered as follows:
\begin{lemme}\label{aRetrouver}
 Let $F$ be a subset of a lattice $T$.
%
%
For all $q,t \in F$ if $q \leq t $, and $x \in T^q$, then $p_t(x)=p_q(x)$ \label{lemEqualProj}
\end{lemme}
\begin{myproof}
 By definition $p_t(x)$ (resp. $p_q(x)$)  is the greatest element of $F^t \cap T_x$ (resp.  $F^q \cap T_x$). As $F^t \subseteq F^q$, we have also that $F^t  \cap T_x \subseteq F^q\cap T_x$.
As both sets have greatest elements,  the greatest element of $F^t \cap T_x$ is also the  greatest element of $F^q\cap T_x$.
\end{myproof}

Theorem \ref{thFondConf} gave us a general condition to obtain closure operators on confluences. We see hereunder that when considering subconfluences of a lattice, we may generalize Proposition \ref{pRondfA} that gave a way to obtain closure operators on abstractions of a lattice from closure operators on the lattice:

\begin{prop}[See \ref{Preuves}]\label{pRondfF}
Let $F$ be a subconfluence of a lattice $T$, $f$ be a closure on $T$, and $\{p_t\}$  the family of interior operators associated to $F$, then $f_F$ defined  as $f_F(t)=p_t\circ f(t)$ is a closure operator  on $F$.

\end{prop}

This is a nice result that allows to obtain closures on subconfluences of a lattice. As subconfluences generalize abstractions this will be the basis to extend FCA further than abstract concept lattices by considering either in the intensional or in the extensional side subconfluences rather than lattices, thus obtaining closure operators whose ranges are  confluences  and will result in \emph{ concept  confluences} generalizing concept lattices.  
 In the next sections we show how using results related to Proposition \ref{pRondfF} we may extend FCA. 
 

\section{From Galois lattices to Galois confluences} \label{FromGaloisLattice}
Formal Concept Analysis deals  with an intensional space, a lattice $L$ describing patterns, and  an extensional  space, a lattice  $X$ representing object subsets in which the patterns occur. 

\subsection{Galois lattices and abstract Galois lattices}
We first need to define the nature of the connection between these two spaces:
 \begin{defi}
 Given two ordered sets $X$ and $L$ a Galois connection between $X$ and $L$ is a pair of mappings $(\mathrm{i}: X \rightarrow L, ~\mathrm{e}:L\rightarrow X)$ such that:
 \begin{itemize}
 \item  For any $x$ and $x'$ in $X$, we have that  $x \leq x' \textit{ implies }  \mathrm{i}(x) \geq \mathrm{i}(x')$  
  \item For any $l$ and $l'$ in $L$ , we have that $l \leq l' \textit{ implies }  \mathrm{e}(l) \geq \mathrm{e}(l')$
  \item  For any $x$ in $X$ and $l$ in $F$ we have that $\mathrm{e}\circ \mathrm{i}(x) \geq x$ and  $\mathrm{i}\circ \mathrm{e}(l) \geq l$
 \end{itemize}
 \end{defi}
 
 The following proposition summarizes the main characteristics of the Galois connection whenever $X$ and $L$ are lattices. As a matter of fact only the third  one, depends on this supplementary condition. 
\begin{prop}\label{propConGal} Let  $(\mathrm{i},\mathrm{e})$ be a  Galois connection between two lattices $X$ and $L$
\begin{enumerate}
\item  $f=\mathrm{i} \circ \mathrm{e}$ is a closure operator on $L$ with $f[L]=\mathrm{i}[X]$  as closure subset.
\item $h=  \mathrm{e} \circ \mathrm{i} $ is a closure operator on $X$ with
 $h[X]= \mathrm{e}[L]$ as closure subset.
\item $h[X]$  and $f[L]$ are two anti-isomorphic lattices and the $(x,l)$ pairs where  $l=\mathrm{i}(x)$ and $x=\mathrm{e}(l)$ form a lattice $G$, ordered following $X$, isomorphic to $h[X]$, called the  Galois lattice of $(\mathrm{i},\mathrm{e})$.
\item Let $\equiv_{\mathrm{e}}$ and  $\equiv_{\mathrm{i}}$ the equivalence relations defined respectively as $l \equiv_{\mathrm{e}} l'$ iff $\mathrm{e}(l)=\mathrm{e}(l') $ and  $x \equiv_{\mathrm{i}} x'$ iff $\mathrm{i}(x)=\mathrm{i}(x') $, then $\mathrm{i}(\mathrm{e}(l))$ and $\mathrm{e}(\mathrm{i}(x))$ are  the maxima of the equivalence classes to which belong respectively $l\in L$ and $x\in X$.
\end{enumerate}

\end{prop} 

 Proposition \ref{pRondfA} tells that composing an interior operator with a closure operator results in a closure operator. As a matter of fact applying  an interior operator  to one of the ordered set involved in a Galois connection  results in a new Galois connection:
  \begin{prop}[See \cite{pernelle2002oe}]\label{propAbstLat0}
Let  $X$ and  $L$  be ordered sets,  $(\mathrm{i},\mathrm{e})$ be a Galois connection on $(X,L)$,
 $p$ be an interior operator on $X$ (resp. on $L$) , and $A=p[X]$ (resp. $A=p[L]$), we have that  
 $(\mathrm{i}, p\circ \mathrm{e} )$ $($resp. $(p\circ \mathrm{i},  \mathrm{e}) )$ is a Galois connection on $(A,L)$ $($resp. $(X,A))$.

\end{prop}


The property is symmetrical, i.e. the interior operator  may be composed with either maps $\mathrm{e}$
or $\mathrm{i}$ or both as we may first compose one map with an interior operator, thus obtaining a new Galois connection, and then an interior operator to the other map, also resulting in the  Galois connection  $(p_e \circ \mathrm{e}, p_i \circ \mathrm{i})$. 

 Furthermore, as applying an interior operator to a lattice $X$ results in a lattice, precisely an abstraction of $X$, the resulting Galois connection define a new Galois lattice. 
We have then, following  Proposition \ref{propConGal}, when $A=p[L]$  a closure operator  $p \circ \mathrm{i} \circ \mathrm{e}$ on $A$ with closed elements $p \circ \mathrm{i}[X]$, a closure operator  $ \mathrm{e} \circ p \circ \mathrm{i} $ on $X$ with closed elements $\mathrm{e}[A]$,  and a Galois lattice $G$ of pairs $(x,l)$  of closed elements s.t. $l=p \circ \mathrm{i}(x)$ and $x= \mathrm{e}(l)$.  In this case, $G$ is called an \emph{abstract Galois lattice}.

\subsection{Galois confluences}\label{GaloisPreconfluences}

Following Proposition \ref{pRondfF} we may build a closure operator on a subconfluence $F$ of $T$  by considering the family of interiors operators $\{p_t\}$ associated to $F$, i.e. equivalently the family of abstractions  $p_t[T^t]$. We investigate here what happens when considering a Galois connection between two lattices $X$ and $L$ and reducing one of the lattice to a subconfluence of it.  In what follows we consider, with no loss of generality, subconfluences of the lattice $L$, the results about subconfluences of $X$ being just a rewriting of the results regarding subconfluences of  $L$. 

Our first remark is that since $i \circ e$ is a closure operator on $L$ and $F$ is a subconfluence of $L$ we have the following corollary of Proposition \ref{pRondfF}:

\begin{theo}\label{extensionalPreconfluence}
Let  $(\mathrm{i},\mathrm{e})$ be a Galois connection on the pair of lattices $(X,L)$,  $F$ be a subconfluence of  $L$, then  $f: F \rightarrow F$ defined as 

$$ \forall  t \in F, f(t)= p_t\circ \mathrm{i} \circ \mathrm{e}(t) $$

is a closure operator on $F$.
\end{theo}

We will further define a Galois confluence isomorphic to $f[F]$ by considering the pairs $(x=\mathrm{e}(t),t)$ where $t$ is closed, i.e. $t=f(t)$. 
Hereunder we investigate how this confluence  may be obtained  as a set of Galois lattices.
As a subconfluence of $L$ is a set of abstractions of $L^t$, we first note hereunder that  we may define a set of Galois connections, as far as we consider for each $L^t$ only elements $x$ of $X$ such that  $ \mathrm{i}(x) \in L^t$. In what follows we still note $\mathrm{i}$ and $\mathrm{e}$ their restrictions to part of their domain.

\begin{prop}\label{gcOnXe}
Let $t$ be an element of $L$, then the restrictions of $\mathrm{i}$ and $\mathrm{e}$ to respectively $X_{\mathrm{e}(t)} $ and $L^t$ are such that 
$$(\mathrm{i}, \mathrm{e}) \mbox{  define a Galois connection on } ( X_{\mathrm{e}(t)} , L^t)$$

\end{prop}

\begin{myproof}
As $\mathrm{i}$ and $\mathrm{e}$ define a Galois connection on $X,L$ we know that they are anti monotonic and therefore:
\begin{itemize}
\item
	For any $y \in L^t$, $\mathrm{e}(y) \leq e(t)$ i.e.  $\mathrm{e}(y) \in X_{\mathrm{e}(t)}$, and as a consequence $\mathrm{e}[L^t] \subseteq X_{\mathrm{e}(t)}$
\item 
	For any $x \in X_{e(t)}$, $\mathrm{i}(x) \geq \mathrm{i}(\mathrm{e}(t)) \geq t$ (as $\mathrm{i}\circ \mathrm{e}$ is a closure operator)  and as a consequence $\mathrm{i}(x) \in L^t$, i.e. $\mathrm{i}[X_{\mathrm{e}(t)}] \leq L^t$

\end{itemize}
This means that the restrictions of these functions have domains and co-domains as follows: $\mathrm{e}: L^t\rightarrow X_{\mathrm{e}(t)}$ and $\mathrm{i}: X_{\mathrm{e}(t)} \rightarrow L^t$, and as they inherit  the properties of $(\mathrm{i},\mathrm{e})$ on $(X,L)$, they also define a Galois connection on $(X_{\mathrm{e}(t)}, L^t)$.

\end{myproof}

This  leads to the following corollary of Proposition \ref{gcOnXe}:

\begin{coro}\label{propCoro}
Let  $(\mathrm{i},\mathrm{e})$ be a Galois connection on the pair of lattices $(X,L)$,  $F$ be a subconfluence of  $L$,  and $p_t$ the  interior operator associated to $t$ in $F$, then
\begin{enumerate}
\item $ f_t= p_t \circ \mathrm{i}\circ \mathrm{e} \mbox{ is a closure operator on } F^t$\label{deux}

with  $ f_t[F^t]=p_t\circ \mathrm{i}[X_{\mathrm{e}(t)}]$
\item $ h_t=\mathrm{e} \circ p_t \circ  \mathrm{i} \mbox{ is a closure operator on } X_{\mathrm{e}(t)} $

with $ h_t[X_{\mathrm{e}(t)}]= \mathrm{e}[F^t]$\label{trois}
\end{enumerate}

\end{coro}

$f_t$ and  $h_t$ are called  \emph{local closure operators}.

We have seen that the local closure operators $f_t$ allow to define a closure operator on $F$,   in the following result we show that  by joining the ranges of the local closure operators $h_t$ we obtain the range $e[F]$:

\begin{theo}\label{propFondConfExt}
Let  $(\mathrm{i},\mathrm{e})$ be a Galois connection on the pairs of lattices $(X,L)$,  $F$ be a subconfluence of  $L$, then

$$\mathrm{e}[F]=\bigcup_{t \in F} h_t[X_{\mathrm{e}(t)}] = \bigcup_{m \in \mathrm{min}(F) }h_m[X_{\mathrm{e}(m)}] $$

\end{theo}

\begin{myproof}
Regarding the first equality, from Proposition \ref{propAbstLat0} and considering  $p$ as the identity function, we deduce that the right part of the equality  rewrites as $\bigcup_{t \in F} \mathrm{e}[F^t]$. Furthermore, as $F$ may be rewritten as the union of its up sets, we have that $F=\bigcup_{t \in F}F^t$. By applying the intensional function to both sides we obtain $\mathrm{e}[F]= \mathrm{e}[\bigcup_{t \in F}F^t]$. As $ \mathrm{e}[F]$ is the image of $F$ by $\mathrm{e}$, it is straightforward that 
$ \mathrm{e}[F]=\bigcup_{t \in F}\mathrm{e}[F^t]$.

The second equality states that we only need the closure operators associated to the minimal elements of $F$. From Theorem \ref{propFondConfExt} we have $\mathrm{e}[F]=C=\bigcup_{t \in F) } h_t[X_{e(t)}]$.\\
Let $C'=\bigcup_{m \in \mathrm{min}(F) } h_m[X_{e(m)}]$, as $\mathrm{min}[F] \subseteq F$ we clearly have $C' \subseteq C$. We have then to show that any element in $C$ may be rewritten as an element of $C'$.
Let then $c$ be an element of $C$, this means that there exists $t\in F$ and $x\in X_{\mathrm{e}(t)}$ such that $c=h_t(x)$. First, there necessarily exists $m \in \mathrm{min}[F]$ such that $t \geq m$ and also that because $\mathrm{e}$ is anti monotonic we have $\mathrm{e}(m) \geq \mathrm{e}(t)$ and therefore, whenever $x$ belongs to $X_\mathrm{e}(x)$ it also belongs to $X_\mathrm{e}(m)$. Now, let $z=\mathrm{i}(x)$, recall that  $z \in L^t$ since we have to apply $p_t$ to $z$ to build $h_t(x)$) and we have seen that $t \geq m$. From Lemma \ref{lemEqualProj}, we can then deduce that $p_t(\mathrm{i}(x))=p_m(\mathrm{i}(x))$ and therefore $c=\mathrm{e}\circ p_m \circ \mathrm{i}(x)$ with $x \in X_\mathrm{e(m)}$. This means that $c$ belongs to $C'$.
Overall we have shown that $C'=C=\mathrm{e}[F]$
\end{myproof}

This generalizes Proposition \ref{propAbstLat0}: we  now have that the union of local closed elements of $X$ with respect to a confluence $F$ of  $L$ is the range of $X$ under the  operator $\mathrm{e}$. Again  we only need the set of minimal elements $\mathrm{min}(F)$. 

The confluence $f[F]$ is isomorphic with the set of pairs $P= \{(x,l) | x \in F, l=f(l), x=\mathrm{e}(l)\} = \{(x,l) | l \in F, l=p_t\circ \mathrm{i}(x) \mbox{ for some } t\in F \mbox{ such that } t \leq \mathrm{i}(x)  , x=\mathrm{e}(l)\}$.  This leads to generalize  Galois  lattices and define Galois confluences:
\begin{defi}\label{extGaloisPreConfluence}
The set $P=\{(x,l) | l \in F, l=f(l)), x=\mathrm{e}(l)\}$ is a confluence isomorphic with $f[F]$ and is called the  Galois confluence defined on the confluence $F$ of  $L$ and the lattice $X$ by the maps  $\mathrm{i}: X\rightarrow L$, and  $\mathrm{e}: L\rightarrow X$ 

\end{defi}

This Galois confluence is the union of its up sets, i.e. the Galois lattices  of $(\mathrm{i}, \mathrm{e})$ on  $( X_{\mathrm{e}(t)} , L^t)$.
To summarize, by reducing the lattice involved in a Galois connection to a confluence, we obtain a set of Galois connections, that can be reduced to the set of those associated to minimal elements of the confluence. As guaranteed by  Proposition \ref{pRondfF},  we obtain a closure operator on this confluence but also obtain a characterization of the corresponding Galois confluence as a set of Galois lattices.

\section{Formal Concept Analysis}\label{formalConceptAnalysis}

In standard Formal Concept Analysis, $X$ is $2^O$ where $O$ is the set of objects in which patterns of $L$ may occur while $L$ is $2^I$ where $I$ is a set of attributes also called \emph{items}. The Galois lattice  is then  a \emph{concept lattice} whose elements $(x,l)$, called \emph{concepts}, have  $x$ as their  \emph{extent} and $l$ as their \emph{intent}.  More recently, this formulation has been extended to better cope with wider data representations. In particular, the intensional space $L$ may be any pattern language with a lattice structure, allowing then to represent interval structures or logical conjunction as well as sets of subgraphs  \cite{ferre04,ganter01pattern,Diday2003rm}. Note that, any Galois connection between two lattices may be rewritten as the connection between two powersets
and therefore there is no strict gain in expressive power in the more general setting. However, the direct formulation of the pattern language as a powerset is often useful  and we will use it in what follows. In more recent FCA the pattern language is often called a \emph{pattern structure}.  Additionally, in order to simplify the resulting representation according to an external point of view on data, interior operators has been used to obtain abstract concept lattices. 

\subsection{Concept lattices and abstract concept lattices}

The standard case in which closed patterns are searched for is when the language is a lattice and that closure  of a pattern relies on the occurrences of the pattern in a set of objects. In data mining the set of occurrences is known as the \emph{support set } of the pattern.

\begin{defi}
Let $L$ be an ordered set and $O$ be a set of objects, a relation of occurrence on $L \times O$ is such that  if $ t_1 \geq t_2 $ and $t_1$ occurs in $o$ then $t_2$ occurs in $o$. 

The \emph{support set}  of $t$ in $O$ is defined as $\mathrm{ext}(t)=\{o \in O \mid t \mbox{ occurs in } o \}$.


\end{defi}
This means that $L$ is ordered following a \emph{specificity} order: if $t_1$ is more specific (or less general) than $t_2 $ then the support set of $t_1$ is included into the support set of $t_2$. 
%
%
When $L$ is a lattice, the interesting case is the one in which objects are  described as  elements of $L$ through a mapping  $d: O \rightarrow L$.   Then $o \in ext(t)$ if and only if $d(o) \geq t$, i.e. an object is in the support set of some pattern whenever its description is more specific than the pattern: 
 
\begin{prop}\label{propLat}

Let the pattern language $L$ be a lattice, $O$ be a set of objects, and $d: O \rightarrow L$ the description function, then for any subset $e$ of $O$  

$$\mathrm{int}(e)={\bigwedge_{o \in e }}~d(o)$$
is the greatest pattern more specific or equal than all objects in $e$
and $(\mathrm{int}, \mathrm{ext})$ 
is a Galois connection on $(2^O, T)$.

\end{prop}
$\mathrm{int}(e)$ is called the \emph{intension} of $e$, and the corresponding Galois lattice $G$ is called a concept lattice.

Let us consider the equivalence relation $\equiv_{\mathrm{e}}$ on $L$ associated to support sets equality. The maximal elements  of an equivalence  class associated to some support  are then defined as support-closed. Following Proposition \ref{propConGal}, such a class has a greatest element  that can be obtained from any of its elements $t$  by applying a closure operator:  $f(t)=\mathrm{int}\circ \mathrm{ext} (t)$. The support-closed elements form exactly the closure subset $f[T]$ and each of them represents the class associated to its support. In this case,   $f$ is then denoted as a \emph{support closure} operator.
In the standard  FCA case, the lattice is a powerset $2^X$ of attributes, the description of an object $i$ is the subset of attributes in relation with $i$ and the Galois lattice  formed by   pairs of corresponding closed elements in $2^X$ and $2^O$ ordered following $2^O$ is called  in the FCA community \emph{a concept lattice}\cite{Gan99}. We also use this terminology in our  wider presentation. 

The set of frequent support closed patterns, i.e. the support-closed elements with support greater than or equal to some threshold, 
represents then all the equivalence classes corresponding to frequent supports. Such a class has also minimal elements, called generators. When the patterns belong to $2^X$,  the min-max basis of implication rules\cite{pasquier2005generating} that represents all the implications $t \rightarrow t'$ that hold on $O$, i.e. such that $\mathrm{ext}(t) \subseteq \mathrm{ext}(t')$, is defined as follows:

 $m=\{ g\rightarrow f \backslash g \mid f \mbox{ is a closed pattern }, g \mbox{ is a generator } f \not = g, \mathrm{ext}(g)=\mathrm{ext}(f)\} $
 
  \subsection{Abstract concept lattices }\label{fermesAbstraits}

Proposition \ref{propAbstLat0} allows to define abstract concept lattices, where the interior operator applies to the extensional space $X=2^O$ \cite{Soldano:2011fk} or to the \emph{intensional space} therefore obtaining \emph{projected pattern structures} as proposed in \cite{ganter01pattern} or both \cite{pernelle2002oe}, The general purpose is to reduce the concept lattice sizes, i.e. the number of  concepts, by simplifying the pattern language   or the extensional space.

\section{ Concept confluences as union of concept lattices } \label{confluenceDeGalois}

Concept confluences  follow then from the results of Section \ref{GaloisPreconfluences}. A concept confluence is simply defined as a Galois confluence   associated to the Galois connection $(\mathrm{int},\mathrm{ext})$
As mentioned before, any of the two lattices involved in the Galois connection can be relaxed to be a confluence, leading to either an \emph{intensional concept confluence} when $L$ is reduced to $F \subseteq L$ or to an \emph{extensional  concept confluence} when $X=2^O$ is reduced to $F \subseteq 2^O$. In both cases, reducing the lattice to a confluence generalizes the reduction of the lattice to an abstraction,  which we have seen to be special cases of confluences. However, in the particular case of abstraction both lattices can  be reduced together, stil leading to an abstract concept lattice, while in the wider confluence case, the symmetry is disrupted. As we have seen in Section \ref{GaloisPreconfluences},  a closure operator is defined on the confluence $F\subseteq L$ while  on the  lattice $X$  we  have a set of closure operators whose union of ranges is the range of the map $e$. As we will see in Section  \ref{patternsFermesAbstraits} we still may reduce one of the lattice to be a abstraction, which still is a lattice, while the other is reduced to a confluence, and still obtain a concept  confluence. What happens whenever both lattices are reduced to confluences is beyond the scope of this article, and will be briefly discussed in the conclusion.  Now, in  the formulation in Section \ref{GaloisPreconfluences} the choice to reduce the  lattice $L$ to a confluence was in purpose: in the remaining of the article we reduce this way the intensional  lattice to be a subconfluence of patterns and examine the consequences of this reduction in formal concept analysis. 

In this case, the concept  confluence  is defined following Definition \ref{extGaloisPreConfluence} where $F$ is a subconfluence of $L$, $X=2^O$, $i=\mathrm{int}, e= \mathrm{ext}$. The elements of the concept confluence are therefore pairs of closed elements $(e,c)$ where $c$ is support-closed, i.e. $c= p_c\circ \mathrm{int} \circ \mathrm{ext} (c)$ and $e=\mathrm{ext}(c)$ is the support set of $c$. 

Theorem \ref{extensionalPreconfluence} tells then that when we consider a dataset of objects $O$ described as elements of the pattern lattice $L$ and reduce $L$ to be a subconfluence of $F$ of $L$, we have a closure operator $f$ on $F$ such that $f(t)$ is the greatest pattern, i.e. the most specific one, greater than $t$ and sharing the same support set $\mathrm{ext}(t)$, and therefore we can reduce our set of patterns to maximal ones, according to their support set. Conversely, in order to  guarantee that such a support closure operator exists for any set of objects $O$ described in $L$, a subset of $L$ has to be a subconfluence of $L$:

\begin{prop}\label{boley}

Let $F$ be a subset of the lattice $L$, then 
the  support closure operator on $F$ with respect to any set $O$ whose objects are described as elements of $L$ exists  if and only if  $F$ is a subconfluence of $L$.
\end{prop}
\begin{myproof}
 We only need here to show that whenever $F$ is not a confluence, there is always an object   set $O$ such that the support closure operator does not exist. Recall that $F$ is not a confluence means that there is some $t$ in $F$  and some $x,y$  in $F^t$ such that $x \vee y$ does not belong to $F$ (and so $x \not =y$). Consider then $O=\{x \vee y\}$, we have
  that  both $x$ and $y$ occurs in $x \vee y$ and are maximal among patterns of $F$ that occurs in $x \vee y$ since  $x \vee y$ does not belong to $F$.
  Now, a support closure operator $f$ should then be such that $f(x)=x$ and $f(y)=y$ as they are both maximal patterns that occur in  $x\vee y$. Furthermore, consider as $t$ one of the maximal lower bounds of both $x$ and $y$ in $F$ . Then the support closure of $t$, $f(t)$ should be be either $x$ or $y$ 
 But, whatever is the choice $f$ is then not monotone  and therefore $f$ cannot be a closure operator.    \end{myproof}

In the M. Boley and collaborators work \cite{Boley:2010fk}, a confluent system $(S,X)$ is such that $S \subseteq 2^X$.  Defined this way, $S$  is similar to our definition of a confluence  of $2^X$ except that $\bot=\emptyset$ belongs to $S$ and   $x \cup y$ is only required to belong to $F$ when $ x \supseteq  t$ and $y \supseteq t$  for any $t \not =\emptyset$.   
Proposition \ref{boley} is a straightforward adaptation  of the theorem of \cite{Boley:2010fk} in which confluent systems replaces confluences of $T=2^X$, and which prohibits to have any $x \in X$  common to all objects in $O$.

\begin{example}
 $(S=\{\emptyset, ab, ac\}, X=\{abcd\})$  is a set system following the definition of M. Boley and co-authors, but is not a subconfluence of $2^{abcd}$ because $ab$ and $ac$ are both greater than  $\emptyset$ while $abc$ does not belong to $S$. On the contrary,  $F=\{ab, ac\}$ is a subconfluence of $2^{abcd}$. 
Let $O=\{o\}$ be an object set and  $d(o)=abcd$ the unique object description in $2^{abcd}$.There is no closure operator on $S$ as there are two maximal elements  $S$,  namely $ab$ and $ac$  that have same support set $\{o\}$ as  $\emptyset$: $f(\emptyset)$ cannot be defined.  On $F$ we may define a support closure operator  $F$  and we have that  $f(ab)=ab$ and $f(ac)=ac$.
\end{example}

Recall that according to Lemma \ref{aRetrouver}    to compute the support closure of some $t$ we only need $p_m$ where $m \in \mathrm{min}(F)$ and $m \leq t$ and that whether $t$ is greater than two minimal elements $m$ and $m'$ then $p_m(t)=p_m'(t)$.  
  For instance, in our example of connected subgraphs generated by  edges of some graph, the minimal elements are the edges. As a consequence, connected subgraphs under some edge $e$ simply are obtained by projecting subgraphs containing $e$ on their connected component containing $e$.

\subsection{Implications}

Another  question regards the definition and   construction of an implication basis whose implications have both left part and right part in $F$. An implication $p\rightarrow q$ holds on $F$ whenever $\mathrm{ext}(p) \subseteq \mathrm{ext}(q)$ and a basis of such implications is typically made of implications such that both $p$ and $q$ belong to the same equivalence class i.e. $\mathrm{ext}(p)=\mathrm{ext}(q)$.   Whenever $F$ is a lattice, the nodes of the concept lattice represents these equivalence classes and $q$ is a closed pattern i.e. the greatest element of the class, and therefore we have $p \leq q$. As an example the \emph{min-max basis}  is made of the implications $p\rightarrow q$  where $p\not=q$ and $p$ is a minimal element of the class of $q$ \cite{pasquier2005generating}. Whenever $F$ is a confluence, we have seen that each such equivalence class is associated to several closed patterns $q_1 ... q_m$ each being the greatest element of a subclass. We have then in the basis both implications of the form $p_i \rightarrow q_i$ where $p_i \leq q_i$ and both belong to subclass $i$ together with  implications of the form $p_j \rightarrow q_i$ where $j \not = i$ and therefore $p_j$ and $q_j$ are unordered. We extend the idea of the min-max basis to confluences as follows:

\begin{defi}
Let $F$ be a confluence, and $F(e) = \{ t \in F \mid \mathrm{ext}(t)=e\}$, the min-max basis $B=B_i \cup B_e$ of implications in $F$ is defined as the set

$\{ p \rightarrow q  \mid \mathrm{ext}(p)=\mathrm{ext}(q), p \not = q, p \in \mathrm{min}(F(e)), q\in f[F(e)]$  \}

The \emph{internal} sub basis $B_i$ is made of the implications of the form $p_i \rightarrow q_i$ where $p_i \leq q_i$ and the \emph{external} sub basis $B_e$ is made of the implications of the form $p_j \rightarrow q_i$ where $\{p_j, q_j\}$
 are unordered. \end{defi}

There are other implication basis such as the minimal  Guigue-Duquenne  basis \cite{duquenne86} that can be as well extended to the case of confluences.

\subsection{Example}\label{subsecExample}
We consider here the example displayed in  Figure \ref{fig-confPlusObjPlusPreconf}. We have $F=\{a,b,abc,abd,abcd\}$ and  $O=\{ab, abc, abcd\}$. To compute the closures in $F$ we take advantage of the fact that  $F$ has two minimal elements $a$ and $b$ and that for any $t \geq a$ (resp. $t \geq b$) we can write $f(t)=p_a  \circ \mathrm{int} \circ \mathrm{ext} (t)$ (resp. ($f(t)=p_b  \circ \mathrm{int} \circ \mathrm{ext} (t)$).
We obtain then:
\begin{itemize}
\item $f(a)= p_a  \circ  \mathrm{int} (\{ab, abc, abcd\})=p_a (ab)=a$
\item $f(b)=p_b  \circ \mathrm{int} (\{ab, abc, abcd\}) = p_b (ab)=b$
\item $f(abc)= p_a  \circ  \mathrm{int} (\{abc, abcd\})=p_a (abc)=abc$ (we could have used $p_b$ as $abc \in L^{b}$ with the same result $abc$)
\item $f(abd)=p_a  \circ \mathrm{int} (\{abcd\}) = p_a (abcd)=abcd$ (same remark as above)
\item $f(abcd)=p_a  \circ \mathrm{int} (\{abcd\}) = p_a (abcd)=abcd$ (same remark as above)

\end{itemize}
Note that the confluence $F$ is the union of the two lattices $F^a= \{a, abc, abd, abcd\}$ and $F^b=\{b, abc, abd, abcd\}$. Therefore we have 
$f[F)=\{a, b, abc, abcd\}$ which is a confluence whose minimal elements are $f(a)=a$ and $f(b)=b$.  We have that $f[F]=f[F^A] \cup f[F^b]$ where $f[F^a]$ and $f[F^b]$ are  the sets of  closed patterns from the concept lattices built respectively from $(F^a, O^a)$,  and  from $(F^b, O^b)$.  We have here $f[F^a]=\{a, abc, abcd\}$ and $f[F^b]=\{b, abc, abcd\}$. 

Regarding the min-max implication basis we first consider the set of extensions $\mathrm{ext}[F] = \{e_1=\{ab, abc, abcd\},e_2=\{abc, abcd\},e_3=\{abcd\}\} $ together with the corresponding equivalence classes $F(e_1), F(e_2), F(e_3)$. Each each equivalence class  may contain disjoint subclasses $C_a$ and $C_b$ of elements from $f[F^a]$ and $f[F^b]$. We denote hereunder such an equivalence class by $C_a+C_b$. The closed elements are underlined.
  \begin{itemize}
\item $F(e_1)= \{\underline{a}\} + \{\underline{b}\}$
\item $F(e_2)= \{\underline{abc}\} $
\item $F(e_3)= \{abd, \underline{abcd} \} $

\end{itemize}

Figure \ref{fig-confPlusObjPlusPreconf} displays on the left  the confluence $F$, on the middle we have the object set $O$, and on the right  is represented the confluence $f[F]$ of support closed patterns of $F$.
The min-max implication basis is made of the internal basis  $B_i= \{abc \rightarrow abcd\}$ (this implication holds both in $(F^a, O^a)$ and in   $(F^b, O^b)$) plus the external basis $B_e= \{a \rightarrow b, b \rightarrow a \}$.

\subsection{Abstract closed patterns in confluences}\label{patternsFermesAbstraits}

In this section we consider abstract closed patterns as those obtained in \emph{extensionally abstract concept lattices}, called here \emph{abstract concept lattices} for short, by constraining the space $2^O$ to be an abstraction $A=p_A[2^O]$ where $p_A$  is the interior operator associated to $A$. As stated in Section \ref{fermesAbstraits}, Proposition \ref{propAbstLat0} leads to a Galois connection $(\mathrm{int}, p_A\circ \mathrm{ext})$,  the abstract extension $\mathrm{ext}_A(t)$ of some pattern $t$ is the union of the elements of $A$ contained in its (standard) extension, i.e.  $\mathrm{ext}_A=p_A\circ  \mathrm{ext}$.  It is then straighforward   that abstract concept confluences are simply the  Galois confluences of Definition \ref{extGaloisPreConfluence} derived from this new Galois connection:
\begin{theo}\label{propAbsConf}
Let $F$ be a subconfluence of a lattice $L$,  $O$ a set  whose objects are described as elements of $L$, $A=p_A(O)$ an abstraction of $A$, then:

Let $p_t$ denote the local description operators on $F$, we have that

$f_A(t)=p_t \circ \mathrm{int} \circ p_A  \circ \mathrm{ext}(t) $,  where $(\mathrm{int}, p_A\circ  \mathrm{ext})$ is a   Galois connection on $(L,A)$, is a support  closure operator on  $F$ with respect to $A$ and $f_A[F]$ is a confluence.

\end{theo}

We continue here the example of section \ref{subsecExample} by using the abstraction \\
$A=\{\{o_1, o_2\}, o_1, o_3\}\}= \{\{ab, abc\}, \{ab, abcd\}\}$. Recall that  $p_A(e)  ={\cup}_{\{a \in A \mid  a \subseteq e\}} a$. We obtain then:
\begin{itemize}
\item $f_A(a)= p_a  \circ  \mathrm{int} \circ p_A(\{o_1, o_2, o_3\})=a$ as $p_A(\{o_1, o_2, o_3\})=\{o_1, o_2, o_3\}=\{ab, abc, abcd\}$
\item $f_A(b)=p_b  \circ \mathrm{int} \circ p_A(\{o_1, o_2, o_3\}) = b$ (same reason as above)
\item $f_A(abc)= p_a  \circ  \mathrm{int} \circ p_A(\{o_2, o_3\})= \top_a= abcd$ as  $p_A(\{o_2, o_3\})= \emptyset$ and therefore  $p_a \circ\mathrm{int}(\emptyset) = p_a (\top_a)=\top_a$ 
\item $f_A(abd)=p_a  \circ \mathrm{int} \circ p_A (\{o_3\}) = \top_a=abcd$ as $p_A( \{o_3\})= \emptyset$  (as above) 
\item $f_A(abcd)=p_a  \circ \mathrm{int} \circ p_A(\{o_3\}) = abcd$  (same  as above)

\end{itemize}

\begin{figure}[!htbp]

\begin{center}

\includegraphics[width=3.75in]{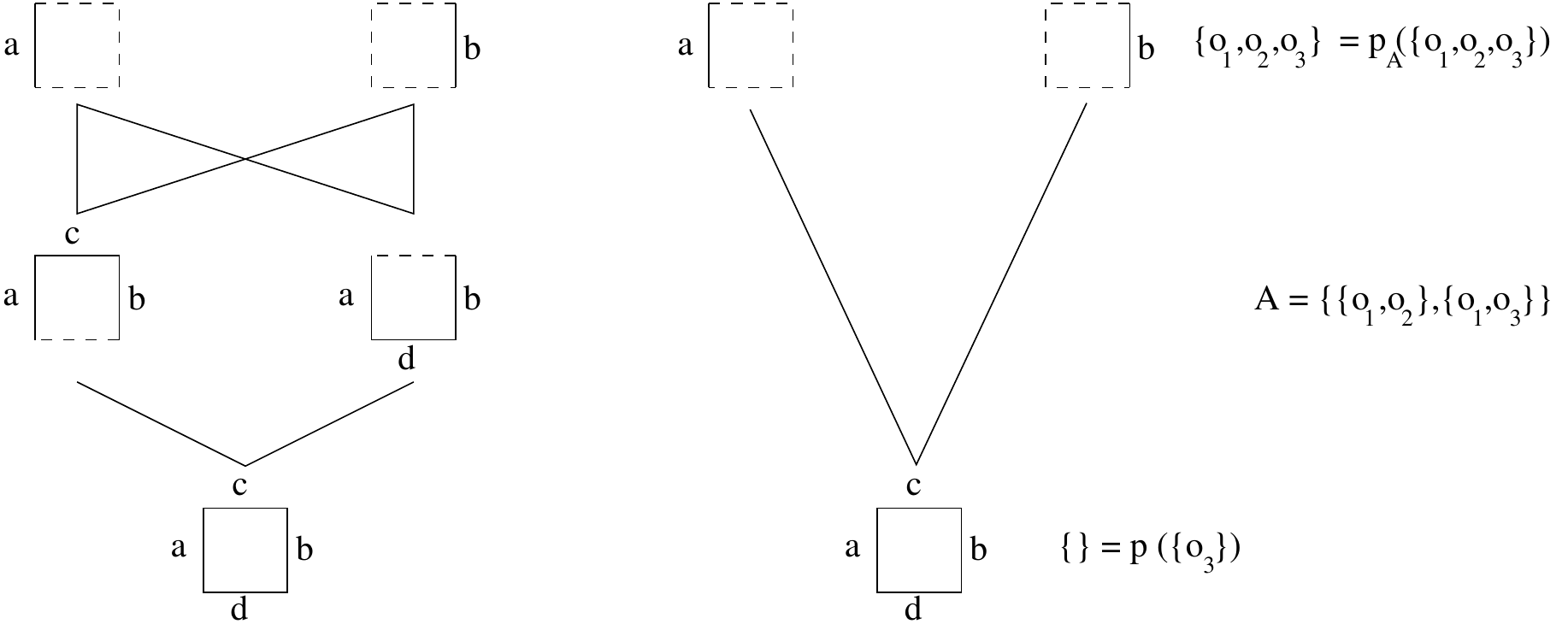}
\caption{ Diagram of the  abstract support closed connected subgraphs confluence $f_A[F]$ (on the left part of the figure) with respect to the abstraction $A=\{\{o_1, o_2\},\{ o_1, o_3\}\}$  of $O$. The support closed element $abc$ of $f[F]$ as been projected to the maximal element of $F$, $abcd$, because its extension $\{abc, abcd\}$ is projected on $\emptyset$ as no element of $A$ is included in $\{abc, abcd\}$.}
\label{fig-confPlusAbstPrec}
\end{center}

\end{figure}

$F$ is represented on the left of Figure  \ref{fig-confPlusAbstPrec}. The corresponding  abstract support closure confluence $f_A[F]$ is displayed on the right of the figure. What happens here, is that there are only two possible support sets  as $\mathrm{ext_A}[F]=\{ \emptyset,  O \}$. As a result the two minimal elements of $f_A[F]$ share the same abstract support set  $O$ whereas the unique maximal element $\top_a=\top_b= abcd$ have an empty abstract support set. 

\section{Algorithmics}\label{algorithmique}

An algorithm  listing the closed elements  of a strongly accessible  set system  $(F, S)$ such that  $\emptyset \in F$   has been proposed  in \cite{Boley2010fk}.  A set system is made of a family $F \subseteq 2^X$ of subsets of $S$, but in what follows we report the  strong accessibility definition of  set systems in \cite{Boley2010fk} by directly referring   to the family $F$.
\begin{defi}
  A family $F \subseteq 2^S$ is called strongly accessible if for any pair of elements $t_1,t_2$ of $F$ such that  $t_1 \subseteq t_2$ there exists a path    $t_1, t_1\cup\{x_1\}, ... , t_1\cup\{x_1, \dots x_k\}=t_2 $ whose elements all belong to $F$.
\end{defi}

The definition of a  confluent set system $(F,S)$ by M. Boley and co-authors \cite{Boley2010fk} is close to our definition of a confluent family $F \subseteq 2^S$ but differs in that no constraint is required on $\emptyset$ that   can then freely belong to any confluent set system. The consequence is that in their analog of Proposition \ref{boley}  it is required that no element of $S$ is shared by all object descriptions in $O$, thus  ensuring that $\emptyset$ is a support closed element and that the support closure operator exists. When applied to listing the  elements of such a strongly accessible confluent set system, including $\emptyset$,  which are support closed with respect to a set of objects, the  algorithm is shown  to output closed elements in a delay which  is polynomial  in the size of the multiset $D=\{ d(o_i) \subseteq X \mid o_i \in O\}$.
This algorithm, adapted to  muticore systems, has been implemented in  PARAMINER \cite{Negrevergne2013qf}. 

Strongly accessible confluences of $2^S$, as defined here,  in general have  several minimal elements and does not contain $\emptyset$.  Adapting the Boley's algorithm to  strongly accessible confluences of $2^S$ requires to start from these minimal elements. We present hereunder an adaptation following the presentation of the original algorithm in \cite{Negrevergne2013qf}.  This allows, for instance to search for support closed elements in the family $F$ of the connected vertex subsets of some graph with size at least $k$. Note that  adding the $emptyset$ element we do  obtain a confluent set system but not a strongly accessible one:  in such a confluent set system, there is no way to reach $\emptyset$ from a vertex susbets of size at least $k$. 

PARAMINER performs a depth-first search, adding an atom  $e$ to a closed pattern $P$, checking whether     $P\cup \{e\}$ is in  $F$, then computing the closure in $F$ of this pattern. PARAMINER uses an exclusion list  $\emph{EL}$ made of atoms of $S$: the algorithm enumerates first the closed elements of $F$ containing the first atom $e_1$, then $e_1$ is put in $\emph{EL}$ and the closed elements of $F$ not containing $e_1$ but containing $e_2$ are enumerated, $e_2$ is put in $EL$ and so on. We add an exclusion list  $\emph{ELM}$ made of minimal elements in $F$: we enumerate first the closed elements  containing the minimal element $m_1$ then those not containing  $m_1$ but another minimal element  $m_2$ and so on. The exclusion list  $\emph{ELM}$ allows to avoid enumerating  twice the same closed pattern starting from two different minimal elements. For this purpose, we use the  function \emph{not\_include\_any\_of(P, ELM )} that returns TRUE  whenever $P$ is included in no element of  \emph{ELM}.  

The function  \emph{Clo} computes the  closure in  $F$ of any pattern $P$  by first computing the support closure of $P$, $\mathrm{int}\circ \mathrm{ext} (P) $ in $2^S$ and then applying the interior operator $p_m$ corresponding to the current minimal element $m$.  
The algorithm  may enumerate  abstract closed patterns in confluences as defined Section \ref{patternsFermesAbstraits}: we just need to generalize the closure function   \emph{Clo} in order to compute the abstract support  closure   $p_m \circ \mathrm{int}\circ  p_A \circ\mathrm{ext} (P) $.   
\begin{tabbing}
   $$ $$ \= $\; \;$ \sc{\; \;}\=\sc{\; \;}\=\sc{\; \;}\= $\;\;$ \sc 
{\; \;}\=\sc{\; \;}\=\sc{\; \;} \= $\; \;$ $\; \;$ \kill

\> {\bf Algorithm 1 }\\
\>{\bf Requires:}	  \\
\> A set  $S$, a strongly accessible confluence $F$ of $2^S$,\\
\> a set $O$  of objects $o$ each described as $d(o)$ in   $2^S$ \\

\>{\bf Ensures:}   \\
\> Ouput the  abstract support closed patterns in $F$ w.r.t. $O$,\\

\> 1: \>  $ELM \leftarrow \emptyset $ \\
\> 2: \> {\bf for all}  $m \in \mathrm{min}(F)$ 
{\bf do}\\
\> 3: \>\> $P \leftarrow \mathit{Clo}(m)$\\
\> 4:\>\> {\bf if} $\mathit{ not\_include\_any\_of}(P, \mathit{ELM}) $ {\bf then}\\
\>5: \>\>\>  $\mathit{enum\_clo} (P,ELM,\emptyset) $ \\
\>6: \>\>\> $\mathit{ELM}\leftarrow \mathit{ELM}\cup \{m\}$  \\
\>7:\>\> {\bf end if}\\

\>8:\>{\bf end for} \\
\> \\
\>  {\bf  Function} $\mathit{enum\_clo}(P, \mathit{ELM},\mathit{EL})$\\
\>{\bf Requires:}	  \\
\>A closed pattern  $P$, an exclusion list   $\mathrm{ELM}$ of minimal elements of $F$,  \\
\> an exclusion list $\mathrm{EL}$ of elements of $S$ \\
\>{\bf Ensures:}   \\
\> Output the closed patterns with $P$ as an ancestor in the enumeration tree     \\
\>1:\> $\mathit{Ouput}  P$ \\
\> 2:\> {\bf for all}  $e $ s.t. $P\cup \{e\} \in F$ {\bf do}\\

\> 3:\>\> $Q \leftarrow \mathit{Clo}(P\cup \{e\})$\\
\> 4:\>\> {\bf if} $ \mathit{ not\_include\_any\_of}(Q, \mathit{ELM} )$ and   $Q \cap  \mathit{EL} = \emptyset$
{\bf then}\\
\>5:\>\>\>  $\mathit{enum\_clo} (Q,ELM,EL) $ \\
\>6:\>\>\> $\mathrm{EL}\leftarrow \mathit{EL}\cup \{e\}$  \\
\>7:\>\> {\bf end if}\\

\>11:\>{\bf end for} \\
\> \\
\> {\bf  Function}  $ \mathit{Clo}(P)$\\
\> {\bf Requires:} A pattern  $P \in F$, an object set $O$\\
\> {\bf Ensures:} Returns the (abstract) support closed pattern  containing   $P$ \\
\> \> and which has the same (abstract) support\\ 
\> 1: \>$ Q \leftarrow \cap_{o \in  p_A\circ \mathrm{ext}(P)} d(o)$ \emph{/*} Closure of $P$ in $2^S$\emph{*/} \\ 
\> 2: \>$m$ $\leftarrow$ an element of $\mathrm{min}(F)$ s.t.. $m \subseteq P$\\
\> 3: \> {\bf return}  $p_m(Q)$  \emph{/*}Closure of $P$ in $F$\emph{*/} \\ 
 \\
\end{tabbing}

This algorithm comes down to the original algorithm  whenever minimal elements are items, except that  the empty set  closure is not computed and that there is then no reason to distinguish the two exclusion lists:  \textit{ not\_include\_any\_of}(Q, \textit{ELM}) is then equivalent to  $Q \cap  \mathit{ELM} = \emptyset$.  Note that, differenty from the  generic algorithm PARAMINER there is a clear distinction between the  family $F$ that has to be confluent and strongly accessible and the selection on closed elements brought by the extensional abstraction. When simply interested in \emph{frequent} closed patterns, i.e. patterns whose support  $\mathrm{ext}(P)|$ is greater than some threshold $s$, the  abstract support set is $\emptyset$ whenever  $|\mathrm{ext}(P)| < s$  and else is unchanged. When the abstract support set is $\emptyset$, the closure operator on $2^S$ returns $S$ and therefore $\mathit{Clo(P,D)}$ returns $\top_P$
 which is the same as $\top_m$ whenever $m \subseteq P$. Note that in this case there is no recursive call as  $\top_m$ cannot be augmented. The only closed patterns for which the  minimal frequency support constraint is not guaranteed is then precisely the $\{\top_m\}$ patterns. 
Outputting the elements $(c,e)$ of the associated concept confluence is straightforward, though ordering them in a confluence needs a further post-processing. 

Detailed proof of correctness and performance analysis of our Algorithm 1 are out of the scope of the present work. However, it is straightforward that correctness relies on correctness of Algorithm 1 in \cite{Boley2010fk} that output closed element in a strongly accessible family containing $\emptyset$. Note that each of the subproblems solved in the calls line 5 of our Algorithm 1 comes down to listing closed elements in a strongly accessible family of $2^{S\setminus m}$ in the very same way as Algorithm 1 in \cite{Boley2010fk}. 
Our Algorithm 1 lists then first  at the first iteration the closed elements of $F^{m1}$ then the closed elements in $F^{m_2}$ that do not belong to $F^{m_1}$ , then those  of  $F^{m_3}$ that do not belong neither to   $F^{m_2}$ nor to $F^{m_3}$ and so on.

\begin{example}
We consider $S=abcde$ and $F = \{ab, ac, abc, abd, acd, abcd\}$ with $\mathrm{min}(F)=\{ab, ac\}$. We also consider an object set $O$ with descriptions $d[O]=\{abde, abcd, acd\}$ and no extensional abstraction, i.e. $p_A= Id_{2^O}$.  The closure operator on $F$  is then s.t. $f(ab)=abd$, $f(ac)=acd$, $f(abc)=abcd$ which are the only support closed elements. Algorithm 1 lists the corresponding closed patterns as follows:

\begin{itemize} 
\item $f(ab)=abd$
\item $\mathit{enum\_clo}(abd, \emptyset, \emptyset)$ 
	\begin{itemize}
	\item $ \rightarrow abd$ 
	\item $\mathit{enum\_clo}(abcd, \emptyset, \emptyset)$
		\begin{itemize}
		\item $ \rightarrow$ abcd 
		\end{itemize}
	\end{itemize}
\item $f(ac)=acd$
\item $\mathit{enum\_clo}(acd, \{ab\}, \emptyset)$	
	\begin{itemize}
	\item $ \rightarrow acd$ 
	\item $ \mathit{ not\_include\_any\_of}(abcd, \{ab\})$ fails as $abcd \supseteq ab$
	\end{itemize}
\item End
\end{itemize}

\end{example}
  
\section{Discussion}\label{discussion}
We have extended in this article the  well-known equivalence  between subsets closed under meet of  a (finite) lattice $T$ and  closure operators on $T$, by weakening the lattice  to be a confluence $F$.
Relaxing lattices to be confluences is a path to introduce such connectivity constraints inside formal concept analysis and closed pattern mining. We obtain  a first nice result:
\begin{itemize}
\item Closure operators on a confluence $F$ are equivalent to subsets $C$ of $F$  closed under local meet, i.e. subsets such that for any minimal element $m$, the part of $C$ greater than or equal to $m$ is closed under the meet operator of $F^m$. As in the lattice case, we have that $f[F]=C$ and that the closure  operators preserve the structure : $f[F]$  is a confluence.

\end{itemize}

Formal concept analysis deals with how some pattern lattice is related to the powerset of a set of objects described as elements of the pattern lattice via a Galois connection. 
 With the purpose of extending formal concept analysis in the direction of simplifying the resulting representation,  it was previously shown that  by applying an interior operator to one of (or both)  the lattices related by a Galois connection, we still obtain two lattices, a Galois connection between them, and an abstract concept lattice smaller than the original concept lattice. 
 As we have investigated  closure operators on  confluences and shown that they preserve the confluence structure, the question arose then of relaxing the abstraction structure to a confluence included in the original lattice. We defined this way a confluence  $F$ of a (host) lattice $T$ as a subset of  $T$  such that each up-set $F^t$ rooted in any element $t$ of $T$ is closed under the join operator $\lor$  of $T$. An abstraction of $T$ is then a subconfluence of $T$ with a minimum. We obtain then the following result:
\begin{itemize}

 \item  Starting from Galois connection between two lattice $L$ and $X$ and restricting one of them, say $L$, to one of its  confluence $F$ we obtain a set of Galois connections between the up-sets of $F$ and $X$, a closure operator $f$ on $F$ and a set of closure operators $h_m$ on $X$.  We also obtain a set of pairs $(e(t),t)$ forming a confluence called a Galois confluence. 

\end{itemize}

Wa may observe that symmetry is disrupted: a closure operator is defined only on  the subconfluence of $L$  while in $X$ we rather have a family of closure operators. As a consequence  all pairs in the Galois confluence have different closed elements of $F$ but two pairs may share a same  element $e(t)$, which is closed w.r.t. two different  closure operators on $X$. When applied to formal concept analysis we have then two directions: either we consider an intensional  subconfluence, i.e. a subconfluence  of the pattern lattice or we consider an extensional  subconfluence , i.e. a subconfluence of the powerset of objects. Section \ref{confluenceDeGalois} investigates the  concept confluences resulting from the former choice.  
There have been recently  promising investigations on the other direction, in which case the powerset of objects may be reduced to connected subgraphs of some graph whose vertices are the objects under investigation and have as  labels their description in the pattern language\cite{Soldano:2015ac,Soldano:2015fk}.


Coming back to the intensional case, the connected vertex subset of Example \ref{connectedEdgeSubsets} should not mislead us about the kind of data analysis problems that can be addressed: general graph mining, in which  matching two subgraphs    leads in general to various  maximal lower bounds (see for instance \cite{Douar2015kq}),   cannot be addressed by defining subconfluences. Formal concept analysis have however  been successfully adapted to  such cases by first  considering  as  intensional space a powerset of subgraphs $2^S$ and then extract maximal subgraphs from them \cite{Kuznetsov2005vn}. There is then a, possibly high, price to pay regarding the number of closed patterns to consider.

The general idea in using subconfluences of pattern languages is to restrict the language w.r.t. some \emph{a priori} bias regarding what relates the primitive elements of the language. Example \ref{connectedEdgeSubsets} displays a subconfluence of $2^E$, where $E$ is the edge set of some graph. This allows to consider relational graphs, as gene interaction graphs, where an object is therefore described as an edge subset, i.e. a set of interactions that occur in this object \cite{Negrevergne2013qf,Yan:2005aa}.   We consider hereunder the case of connected conjunctions by considering connected subgraphs induced by vertex subsets of a graph $G=(S,E)$.
 
 \begin{example}
Let  $X=\{P(a,b), Q(b), P(b,c), Q(c), P(c,d), Q(d)\}$ be a set of ground predicates  and  $G=(S,E)$ be the graph obtained when considering that an edge connect two predicates whenever they share an argument: $(P(a,b), Q(b))$ is an edge but  $(P(a,b), Q(c))$ is not. Consider then  the connected subsets of predicates, representing connected conjunction of positive literals: $(P(a,b) Q(b)P(b,c)$ is connected but $P(a,b)Q(d)P(b,c)$ is not. 
By composing the resulting subconfluence of $2^S$ with an abstraction  (see \cite{Soldano:2015ac}) we may better define the targeted pattern language. For instance, by requiring that in such a pattern, considered as a subgraph, all vertices have degree at least 2, we select  $(P(a,b)P(b,c)P(c,a))$ but not $(P(a,b)P(b,c)P(c,d))$.
\end{example}

In the following case connectivity expresses sequentiality:
\begin{example}
Let $S=a_1, a_2, \dots a_n$ be a sequence of  distinct events that does not always all occur  but always occur in the sequence order. A word is a contiguous subsequence  of events $a_i a_{i+1} \dots a_{i+k}$. Consider then a set  $O$ of observations  each made of a subsequence of $S$.  When considering  $2^S$ as the pattern language,  patterns as $a_1,a_3,a_4,a_5$, are made of various words. When considering $S$ as the vertex set of a graph $G$ whose edges relate, for any $i$,  $a_i$ to $a_{i+1}$, we obtain that the set  $W$ of words is the subconfluence of $2^S$ associated to the connected vertex subsets of $G$ and therefore    support closed words are obtained using a closure operator:  given a word $m$, the closed word   $f(m)$ is the largest word including $m$ and occurring in the same observations.  For instance,  $f(a_3,a_4)$ could be $a_2,a_3,a_4,a_5$. It is also possible to introduce $k$-bounded gaps in the words by adding additional edges. For instance $a_1a_3a_4$ is a 2-bounded gaps pattern since for any   $a_ia_j$ in this subsequence we have that $j-i \leq 2$. The set $W_k$ of  $k$-bounded gaps patterns again is a subconfluence of $2^S$ associated to connected vertex subsets of the previous graph to which $a_ia_{i+2}$ edges have been added. In both cases, the algorithm provided in Section \ref{algorithmique} lists the corresponding support closed patterns. 
\end{example}\

Using confluences in formal concept analysis and closed pattern mining has still to be explored in depth. Any pattern language  which has a least general generalization operator  \cite{plotkin71furtherNote,Cohen1992fk} is a candidate to apply connectivity constraint and extract confluences.  From a theoretical point of view there is still investigations to be performed regarding confluences and  lattice confluences as  order structures. In particular our study here is limited to finite orders. As in the case of lattices,   results and definitions on confluences holds in the case of infinite but complete structures, however in the infinite case  minimal elements may not exist and all consequences of Lemma  \ref{lemmeMin}, as second equality in Theorem \ref{propFondConfExt} should be ignored. 

Regarding algorithmics, Algorithm 1 is restricted to strongly accessible confluent families. However, this strong limitation can be overcome. One way for this is  changing the host lattice in such a way that the confluence becomes strongly accessible, which may however result in a large number of atoms. 
 \section{Conclusion}

There is still much   to explore,  in particular regarding the definition  of implication bases and the algorithms to build them. The construction and  visualisation of a concept confluence diagram is also an open problem.  
A major domain of interest for such an extended FCA is its application to analyse data in complex networks. A first work introduced   graph abstractions and their use, in reducing the extensional space to analyze an attributed graph \cite{Soldano2014cr}. More recently, a work on extensional confluences has been presented in order to extract a confluence of subcommunities, associated to local closed patterns and local implications, in an attributed graph\cite{Soldano:2015ac}.  
 We argue that extending FCA in this direction is a way to widen its scope in data analysis,  as  modeling data in a labelled graph  is an active domain. Further work concerns applications to directed graphs and multiplex networks, and in the theoretical side,  the reduction of both intensional and extensional spaces to confluences. An area of application would then be situations in which observations are connected in some way, as for instance experiments sharing some initial conditions, while patterns are made of possibly connected attributes, as for instance gene expressions in an interaction network. 

  \bibliographystyle{elsarticle-num} 
\bibliography{.//papiersRIA}

\appendix
\renewcommand*{\thesection}{\Alph{section}}
\section {Proofs}\label{Preuves}

\begin{myproof}{\bf Proposition  \ref{pRondf}}
\begin{itemize}
\item $p\circ f$ Extensive on $p[L]$?  \\
Let $x \in p[L]$, we have first $f(x) \geq x$ as $f$ extensive,  then $p\circ f(x) \geq p(x)$ as $p$ monotone, finally $p(x)=x$
as $x$ belongs to $p[L]$ and $p$ is idempotent. As a result, $p\circ f (x) \geq x$ 
\item  $p\circ f$ monotone on $p[L]$ ? As both $p$ and $f$ are monotone, obviously $p\circ f$ is monotone
\item $p\circ f$ idempotent  on $p[L]$? 
	\begin{enumerate}
	\item We have $p\circ f(x)\geq x$ as $p\circ f$ is extensive. Applying $f$ then $p$, both monotones, to this inequality, we obtain $p\circ f \circ p \circ f(x) \geq p\circ f(x)$     \label{idemDir}
	\item We have $f(x) \geq p\circ f(x)$ as $p$ is intensive, and $f\circ f(x) \geq  f \circ p\circ f(x)$  as $f$ is monotone. As $f$ is idempotent, this means $f(x) \geq   f \circ p\circ f(x)$, and finally  $p\circ f(x) \geq   p\circ f \circ p\circ f(x)$. \label{idemInv}
	\end{enumerate}
	From \ref{idemDir} and \ref{idemInv} we deduce  $p\circ f \circ p \circ f(x) =p\circ f(x)$

\end{itemize}
\end{myproof}

\begin{myproof} {\bf Theorem \ref{thFondConf}}

\begin{itemize}
\item $\Rightarrow$ 
$C$ is a closure subset of $F$ means that there exists a closure operator $f: F \rightarrow F$ such that $f[F]=C$.

For any $x \in F^t$, we have that $f(x) \in F^t$ (extensivity of $f$) i.e.  $f[F^t] \subseteq F^t$ and therefore we can then define $f_t: F^t \rightarrow F^t$ such that for any $x \in F^t, f_t(x)=f(x)$. It is straightforward that $f_t$ is a closure on $F^t$, since $f$ is a closure on $F$. Now, in one hand,  we have obviously that $f_t[F^t] \subseteq F^t$ and that $f_t[F^t] =f[F^t] \subseteq C$, and therefore $f_t[F^t] \subseteq C \cap F^t$. In an other hand, consider any $x$ in $ C \cap F^t$, as $x \in C$ we have that $x=f(x)$ and therefore $x=f_t(x)$, as a conclusion $C \cap F^t  \subseteq f_t[F^t] $. From these two inequalities we deduce that $f_t[F^t] = C \cap F^t$. This means that $C \cap F^t$, is the range of a closure operator on the lattice $F^t$, and according to  Proposition \ref{proFond} is closed under $\land_t$. As this is true for any $t$ in $F$,  $C$ is closed under local meet in $F$.

\item $\Leftarrow$ 
Let $C$ be a subset of $F$ closed under local meet.  This means that for any $t$, $C\cap F^t$ is a subset of the lattice  $F^t $ which is closed under the meet operator $\land_t$ and therefore is also a lattice. As a result of Proposition \ref{proFond} we have then  that there exists a closure operator $f_t$ on $F^t$  such that for any $x \in F^t$, $f_t(x)={\wedge_t}_{c \in C \cap F^t \cap {F}^x } c$ = $ {\wedge_t}_{c \in C  \cap {F}^x } c$. We will consider then the mapping $f: F \rightarrow F$ such that $f(x)= f_x(x)$. It is straightforward that $f$ is a closure operator.
\begin{itemize}
\item For any $x \in F$, $f(x) = f_x(x)$  and   we have that $x \leq f_x(x) = f(x)$  and therefore $f$ is extensive.
\item For any $x,y \in F$ with $x \leq y$ we have that  $f_x(x) \leq f_x(y)$ and   as  $F^y$ is a sublattice of $F^x$  we have $\land_y=\land_x$ and therefore $f_x(y)={\wedge_x}_{c \in C \cap {F}^y } c$ = $ {\wedge_y}_{c \in C  \cap {F}^y } c$ = $f_y(y)$. We obtain then that $f_x(x)\leq f_y(y)$  i.e. $f(x)\leq f(y)$ and conclude that $f$ is monotone.

\item Let $y=f_x(x)=f(x)$. As $f_x$ is idempotent we have $f_x(x) =f_x(y)$. As above,  $F^y$ is a sublattice of $F^x$  and therefore $f_y(y)=f_x(y)$. We obtain then that $f_x(x)=f_y(y)$  i.e. $f(x)=f(f(x))$ and conclude that $f$ is idempotent
\end{itemize}

We may rewrite $C$ as $C= \cup_{t \in F} C \cap F^t$. For any $t \in F$ we have by definition   $ C \cap F^t = f_t[F^t] =f[F^t]$, and therefore $C= \cup_{t \in F} f[F^t] = f[F] $. As a conclusion $C$ is a closure subset of $F$. 

\end{itemize}

\end{myproof}

\begin{myproof} {\bf Proposition \ref{pRondfF}}
From  Proposition \ref{propEqui}   we know that for any $t\in F$, $F^T=p_t[T^t]$ is the abstraction of the lattice $T^t$ associated to the interior operator $p_t$. Following Proposition \ref{pRondfA} this means that $p_t\circ f$ is a closure operator on $F^t$ we further denote by $f_t$.  $f_F$ is then defined by  $f_F(t)=f_t(t)$.
We obtain  then that $f_F$ is a closure operator on $F$:
\begin{itemize}
\item $f_F$ extensive ? For any $t \in F$, $f_F(t)=  f_t(t) \geq t$ as $f_t$ is extensive on $F^t$ 
\item $f_F$ monotone ? For any $x,y \in F$ and $y \geq x$, we have $ f_x(x) \leq f_x(y)$ as $f_x$ is monotone. Furthermore we have $f_x(y) =p_x\circ f(y) = p_y\circ f(y) =f_y(y)$ following Lemma \ref{lemEqualProj}, and therefore $f_x(x) \leq f_y(y)$ i.e. $f_F(x) \leq f_F(y)$.
\item $f_F$ idempotent ? For any $x\in F$, let $y=f_x(x)$. We have $f_x(x) =f_x(y)$ as $f_x$ is idempotent. But we also know,   again following Lemma \ref{lemEqualProj}, that $f_x(y) =f_y(y)$, and as a result $f_x(x)=f_y(y)$ i.e. $ f_F(x) =f_F(y) =f_F(f_F(x))$. \end{itemize}
\end{myproof}

\end{document}